\documentclass[showpacs,amsmath,amssymb,pre,twocolumn,floatfix]{revtex4}
\usepackage{graphicx}

\begin{document}

\title{Domain magnetization approach to the isothermal critical exponent}

\author{A.-M. Tsopelakou}
\author{G. Margazoglou}
\author{Y. F. Contoyiannis}
\author{P. A. Kalozoumis}
\author{F. K. Diakonos}
\affiliation{Department of Physics, University of Athens, GR-15771 Athens, Greece}

\author{N. G. Fytas}
\affiliation{Applied Mathematics Research Centre, Coventry
University, Coventry CV1 5FB, United Kingdom}

\date{\today}

\begin{abstract}

We propose a method for calculating the isothermal critical
exponent $\delta$ in Ising systems undergoing a second-order phase
transition. It is based on the calculation of the mean
magnetization time series within a small connected domain of a
lattice after equilibrium is reached. At the pseudocritical point,
the magnetization time series attains intermittent characteristics
and the probability density for consecutive values of mean
magnetization within a region around zero becomes a power law.
Typically the size of this region is of the order of the standard
deviation of the magnetization. The emerging power-law exponent is
directly related to the isothermal critical exponent $\delta$
through a simple analytical expression. We employ this method to
calculate with remarkable accuracy the exponent $\delta$ for the
square-lattice Ising model where traditional approaches, like the
constrained effective potential, typically fail to provide
accurate results.

\end{abstract}
 \maketitle

\section{Introduction}
\label{sec:introduction}

The statistical properties of interacting spins on the lattice are
of fundamental importance for our understanding of equilibrium
phase transitions in nature. Among them the most studied candidate
is the Ising model~\cite{lenz,ising} in one, two, and three
spatial dimensions finding application in a wide range of physical
systems ranging from pure and random
ferromagnets~\cite{baxter,binder77,nattermann98,newman99,landau-binder}
to neurons~\cite{hopfield82,amit89} and hadronic
matter~\cite{kogut04}. For $D \geq 2$ spatial dimensions the usual
Ising model for a lattice with infinite size possesses a thermal
phase transition of second order, signaling the spontaneous
symmetry breaking of mirror symmetry and the emergence of a
ferromagnetic (for attractive spin-spin interaction) phase. This
transition is characterized by the presence of a critical
temperature $T_{\rm c}$ at which a scale-free behavior dominates
the underlying phenomenology~\cite{fisher71,privman90,binder92}.
The absence of a characteristic scale at the critical point is
related to the divergence of the correlation length $\xi$ which in
turn leads to the appearance of a variety of power laws with
associated critical exponents. One of the main tasks in the study
of models showing critical behavior is the determination of the
values of these exponents which are then related to the respective
universality classes.

For $D = 2$ the Ising model with next-neighbor interactions is
analytically solvable~\cite{onsager44} while for $D > 2$ one has
to rely on numerical simulations and other approximate type of
solutions for its study, see e.g.~\cite{newman99,landau-binder}
and references therein. Employing numerical methods constrains the
mathematical analysis on lattices with finite size introducing an
additional scale in the problem. This adds a complication in the
development of methods to calculate the critical temperature as
well as the critical exponents, usually solved by the so called
finite-size scaling analysis~\cite{fisher71,privman90,binder92}.
This requires the determination of observables in lattices of
increasing size in order to extrapolate to the infinite size
limit. Coming back to the 2D Ising model, despite of being
analytically solvable, it offers a playground for testing a
variety of simulation algorithms~\cite{newman99} and in fact a
very hard one, since its finite-size scaling analysis possesses
severe peculiarities due to the presence of significant,
non-universal, logarithmic corrections~\cite{ferdinand69,mccoy}.

A dominant feature of the critical state in Ising-like systems,
consequence of the absence of a characteristic length scale, is
the formation of self-similar ordered clusters of spins with
fractal geometry. The associated fractal dimension of the
resulting set of ordered clusters is related to the isothermal
critical exponent $\delta$ which in turn determines the
universality class of the undergoing transition. Thus, the
calculation of the exponent $\delta$ is a very significant task in
the simulations of critical systems. Usually, in order to obtain
the value of the exponent $\delta$ one can either calculate the
fractal dimension of the ordered clusters at $T_{\rm
c}$~\cite{coniglio89} or estimate the so called constrained
effective potential~\cite{fukuda75}. For lattices with finite
sizes the fractal geometry is an approximate property expressed
through self-similar structure between two scales, the constrained
effective potential acquires finite volume corrections and the
critical temperature is replaced by the pseudocritical one, making
the related analysis a highly non-trivial task. Nevertheless the
aforementioned methods work quite well for the case of the 3D
Ising model~\cite{tsypin94}. However they practically fail to give
the right answer in the 2D case since they suffer from the
presence of large finite-size corrections, that are hard to
control.

In the present work we develop a method to calculate the critical
exponent $\delta$ for Ising systems based on properties of the
magnetization time series after thermal equilibrium is reached.
The idea is to consider the mean magnetization time series
obtained by averaging over a connected small subset of the
lattice. The main advantage of considering such small domains
relies on the fact that the number of the microstates determining
their thermodynamic properties decreases exponentially with
decreasing domain size. Thus, for small domains a much better
covering of the available phase space can be achieved. On the
other hand the domains are always open systems and therefore their
thermodynamic properties are quite different from those of the
entire (closed) lattice. Based on the distribution of the mean
magnetization values within such a small domain one observes that
each domain equilibrates at its own effective temperature. In fact
with decreasing size the effective temperature decreases too. The
method of analysis of the domain magnetization time series at
equilibrium makes use of fact that for a closed system the
distribution of waiting times in the neighborhood of the ``false"
vacuum, exhibiting the spontaneous symmetry breaking, is known for
temperatures $T \lesssim T_{\rm c}$. Namely at $T=T_{\rm c}$ the
distribution is a power law with an exponent directly related to
the isothermal critical exponent $\delta$~\cite{Contoyiannis2002},
while for $T < T_{\rm c}$ the power law gets gradually destructed
and exponential tails appear. However, for small to moderate
waiting times the power-law description is still valid and the
connection to $\delta$ holds too~\cite{Contoyiannis2007}.

We use here this information to extract the isothermal critical
exponent $\delta$ for an Ising system. Our basic assumption,
validated by our numerical results, is that the waiting-time
distributions derived in
Refs.~\cite{Contoyiannis2002,Contoyiannis2007} are valid also for
an open Ising spin system in a thermal environment. Thus, we
consider an ensemble consisting of domains of increasing size,
parts of an Ising lattice at the corresponding (pseudo)critical
temperature. For each such domain we calculate the mean
magnetization as a function of time for a long time interval. Time
is measured in single spin flips of the entire lattice. Based on
this time series we calculate the distribution of waiting times in
the neighborhood of $\langle M_{\mathcal{D}_i} \rangle = 0$ for
each domain $\mathcal{D}_i$ and from these distributions we
determine $\delta$. For very small domains the corresponding
effective temperature is also extremely small and the distribution
of the waiting times becomes pure exponential containing no
information for the critical exponent $\delta$. Thus we use a
cutoff on the minimum size of the considered domains. We apply
this approach to the 2D Ising model on the square lattice with
nearest-neighbor interactions and we obtain a highly accurate
estimate for $\delta$. Since the developed method is quite
general, it can in principle be applied to a wide class of spin
lattices in two, but also higher dimensions.

The rest of paper is organized as follows: In
Sec.~\ref{sec:strategy} we present the general strategy and the
method of analysis employed to calculate $\delta$ for arbitrary
spin models of Ising type on the lattice. In
Sec.~\ref{sec:results} we apply the proposed method to the case of
the 2D Ising model on the square lattice and we discuss the
obtained results. Finally, our concluding remarks are presented in
Sec.~\ref{sec:conclusions}.

\section{Strategy and method of analysis}
\label{sec:strategy}

We consider the fluctuations of the magnetization for a spin model
of the typical Ising form at thermal equilibrium defined by the
Hamiltonian~\cite{landau-binder}
\begin{equation}
\mathcal{H} = - \sum_{\langle i,j\rangle} J_{ij} s_i s_j,
\label{eq:Ising}
\end{equation}
where as usual $s_i = \pm 1$ and $\sum_{\langle i,j\rangle}$
indicates summation over nearest neighbors. The condition
$J_{ij}>0$ restricts the following considerations to the
ferromagnetic case, although most of the obtained results are
expected to hold more generally. We are interested for the time
evolution of the magnetization calculated in a connected compact
domain $\mathcal{D}$, subset of the entire lattice. Thus, the
observable we focus on is defined via
\begin{equation}
M_{\mathcal{D}}^{(L_d)}(n) = \sum_{i} s_i(n)
\chi_\mathcal{D}(i)~;~\chi_\mathcal{D}(i)=\left\lbrace
\begin{array}{c} 1~;~i \in \mathcal{D}  \\ 0~;~i \notin
\mathcal{D}\end{array} \right. \label{eq:obser1}
\end{equation}
$L_d$ counts the number of sites in $\mathcal{D}$ and $n$ is the
time unit measured in the minimum time needed for the flip of a
single spin. Clearly $1 \leq \ell \leq N_L$ where $N_L$ is the
total number of sites of the lattice. When $\ell$ approaches $N_L$
then $M_\mathcal{D}^{(L_d)}(n)$ approaches the total magnetization
of the spin lattice $M_L$. An important property of the
magnetization time series is that it carries information of the
approach to criticality, clearly manifested during relaxation in
the effect of critical slowing down~\cite{newman99}. However, the
imprints of criticality are present also to the post-equilibrium
dynamics, as demonstrated in Ref.~\cite{Contoyiannis2002}. It is
the latter information we will use in the present approach. In
fact it is based on the observation that close to the critical
point the magnetization time series becomes {\emph{sticky}} in the
neighborhood of the minimum of the effective potential of the
magnetization which undergoes the spontaneous symmetry breaking.
This behavior resembles intermittent dynamics close to a
bifurcation~\cite{kohyama84}.

\begin{figure}[htbp]
\centerline{\includegraphics[width=8.5 cm]{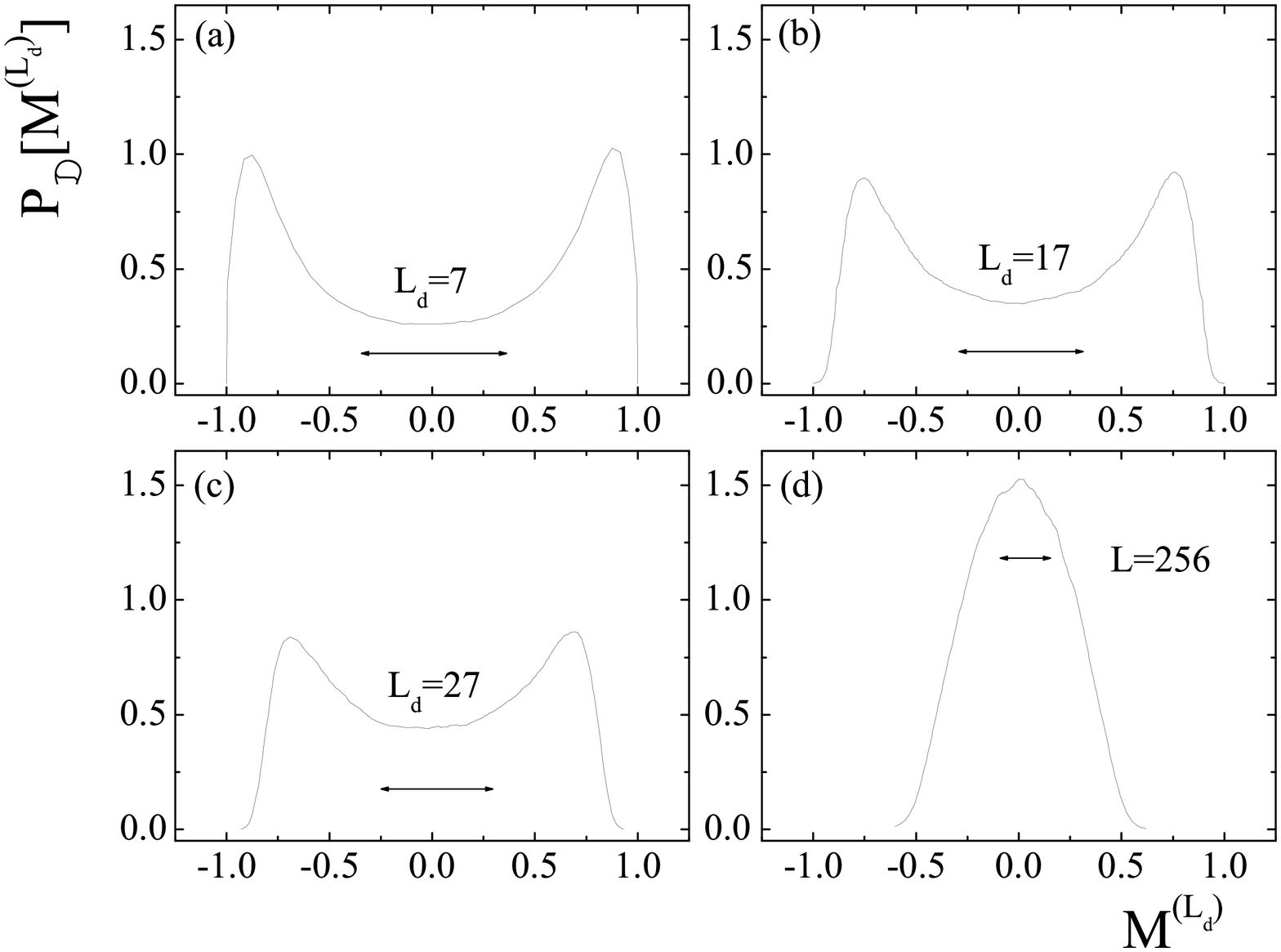}}
\caption{Normalized magnetization probability distribution at the
pseudocritical temperature $T_{c} = 2.29$ calculated in $L_d
\times L_d$ domains of a square Ising lattice with linear size
$L=256$ for various values of $L_d$: (a) $L_d=7$, (b) $L_d=17$,
and (c) $L_d=27$. In panel (d) the corresponding distribution for
the entire lattice is shown. In each plot the arrows indicate the
half-variance region around the mean value $M=0$ used in the
calculation of the waiting-time distribution.} \label{fig:fig1}
\end{figure}

Let us assume that local equilibrium, i.e. equilibrium with the
domain $\mathcal{D}$, is reached before the establishment of
global equilibrium. Then, after reaching global equilibrium, the
domain magnetization time series $M_\mathcal{D}^{(L_d)}(n)$
fluctuates around a stationary mean value. The fluctuations define
a probability distribution $P_\mathcal{D} =
P_\mathcal{D}[M_\mathcal{D}^{(L_d)}]$ characteristic for the
considered domain. For a periodic lattice the exact position of
the domain is irrelevant and the distribution $P_\mathcal{D}$
depends only on the size of $\mathcal{D}$. In fact each domain of
different size is characterized by a different probability
distribution $P_\mathcal{D}$ at equilibrium. This is quite
plausible since pseudocritical temperatures calculated for
lattices of different size increase with decreasing lattice size.
According to this reasoning the pseudocritical temperature of the
entire lattice will be equivalent to an effective temperature
below the pseudocritical one for smaller domains embedded in the
lattice. This can be clearly seen in the distribution
$P_\mathcal{D}$ which is expected to attain the spontaneously
broken form for each such domain.

Concerning criticality the main claim of the present work is that
one can use the time series $M_\mathcal{D}^{(L_d)}(n)$ (leading to
$P_\mathcal{D}$ is the post-equilibrium phase) to decode
information related to the isothermal critical exponent $\delta$.
Based on $M_\mathcal{D}^{(L_d)}(n)$ it is straightforward to
calculate the waiting times $\tau^{(L_d)}$ in the neighborhood of
the false vacuum at $M_\mathcal{D}^{(L_d)}=0$. There is no strict
definition of the term ``neighborhood'' here and the half of the
variance of the distribution $P_\mathcal{D}$ can be used as a
typical neighborhood size.
\begin{figure}[htbp]
\centerline{\includegraphics[width=8.5 cm]{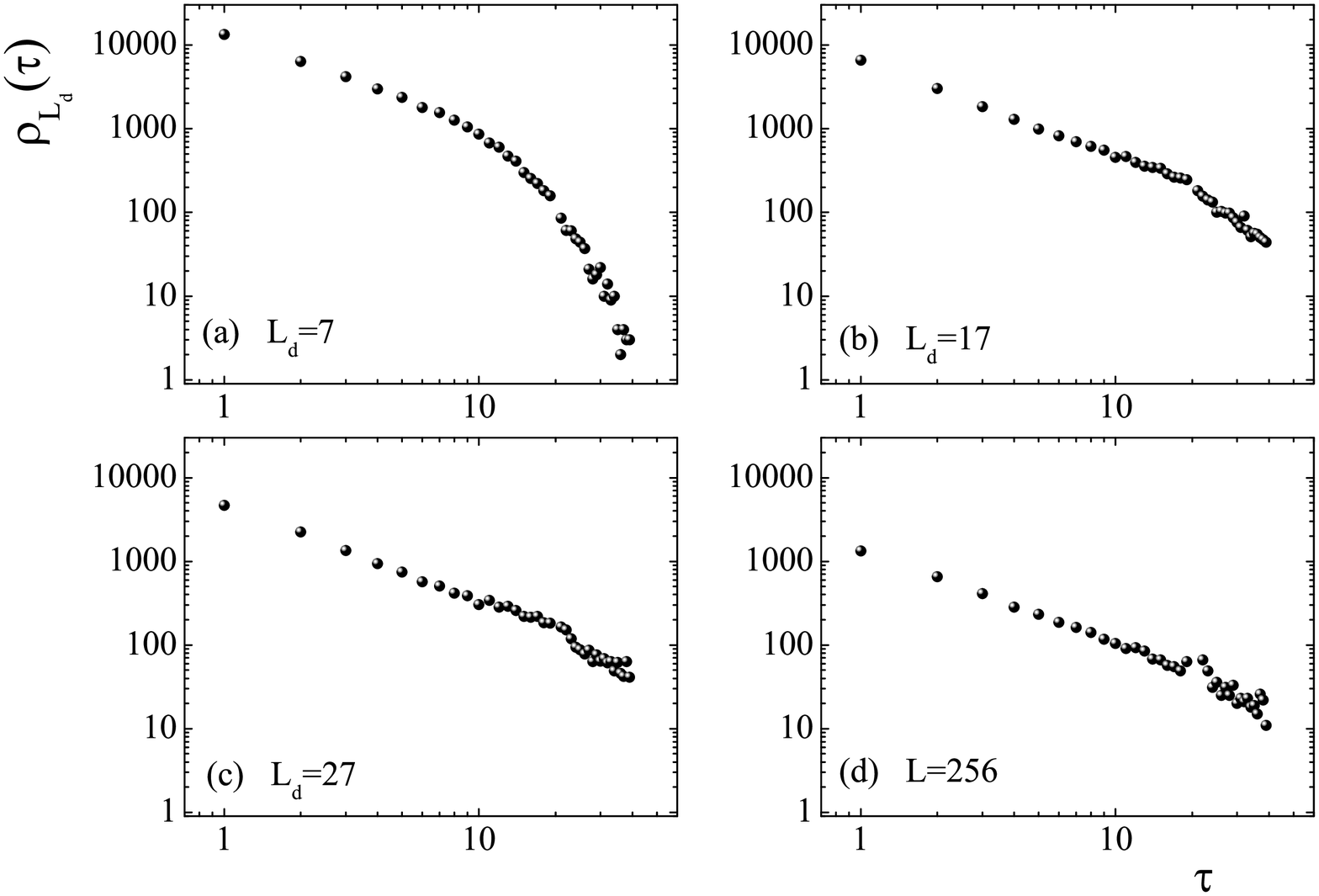}} \caption{The
probability distribution of the waiting times in the neighborhood
of $M=0$ [indicated by the arrows in Fig.~\ref{fig:fig1}(a) - (d)]
at the pseudocritical temperature $T_{c} = 2.29$ for square
domains with linear sizes: (a) $L_d=7$, (b) $L_d=17$, (c)
$L_d=27$, and (d) for the entire lattice. Note that for the panels
(a), (b), and (c), $\rho_{L_{d}}(\tau) \equiv
\rho_{L_{d},bc}(\tau)$, whereas for panel (d) $\rho_{L_{d}}(\tau)
\equiv \rho_{L_{d},c}(\tau)$.} \label{fig:fig2}
\end{figure}
As shown in Ref.~\cite{Contoyiannis2002}, for a finite system of
size $L$ at the pseudocritical temperature $T_{c}$ the
distribution of the waiting times $\rho_{L,c}(\tau)$ has the form
\begin{equation}
\rho_{L,c}(\tau)=c_L \tau^{p_2(L)} e^{-p_3(L) \tau},
\label{eq:lamdiscr}
\end{equation}
where $c_L$ is a normalization factor and $p_2(L)$, $p_3(L)$ are
associated characteristic exponents. For an infinite system
$p_3(\infty) \to 0$ and Eq.~(\ref{eq:lamdiscr}) becomes a pure
power law with $p_2(\infty) \to \frac{\delta + 1}{\delta}$. Thus
the exponent $p_2$ encodes information of the isothermal critical
exponent $\delta$. Interestingly enough, this information is also
encoded in the waiting-time distribution around the false vacuum
just after the spontaneous symmetry
breaking~\cite{Contoyiannis2007}. The corresponding distribution
$\rho_{L,bc}(\tau)$ attains the form
\begin{equation}
\rho_{L,bc}(\tau)=\tilde{C}_L \frac{e^{\tau
(\zeta(L)-1)(r(L)-1)}}{\left [ e^{\tau
(\zeta(L)-1)(r(L)-1)}-1\right ] ^{\frac{\zeta(L)}{\zeta(L)-1}}},
\label{eq:lamdisbcr}
\end{equation}
where $\zeta(L)$ and $r(L)$ are further characteristic exponents.
Note that in both Eqs.~(\ref{eq:lamdiscr}) and
(\ref{eq:lamdisbcr}) above, the under-scripts $c$ and $bc$ refer
to critical and below critical regimes. For an infinite system
$\zeta(\infty)=\delta + 1$ while $r(\infty)$ tends to a
non-universal finite value greater than 1. For small $\tau$ the
density $\rho_{L,bc}$ behaves as a power law $\sim
\tau^{-\frac{\zeta(L)}{\zeta(L)-1}}$ while for large $\tau$ it
decays exponentially $\sim e^{\tau(r-1)}$. Thus
Eq.~(\ref{eq:lamdisbcr}) can be used to estimate $\delta$ for $T
\lesssim T_{c}$. This task will be fulfilled in the next Section.
Using a set of coexisting domains $\mathcal{D}_i$ of different sizes
$L^{(i)}_d$ embedded in an Ising lattice, we will first determine
the waiting-time distribution $\rho_{L^{(i)}_d,bc}(\tau)$ for each
such domain. Then fitting the resulting distribution with a power
law in the small $\tau$ region we will determine the exponent
$\zeta(L)$ and subsequently the isothermal critical exponent
$\delta$. The main assumption here is that each domain can be
considered as an Ising lattice of smaller size than the size of
the lattice it is embedded into. The Ising spins of the entire
lattice, not belonging to the considered domain, form a thermal
environment for this domain at a temperature which depends on the
domain size $L^{(i)}_d$. There is a good reason for adopting this
scenario: as it will be demonstrated in the next Section the
properties of small domains are obtained with much higher accuracy
than those of the larger domains probably due to an exponentially
better covering of the corresponding microstate space. This leads
to a robust estimation of the isothermal exponent $\delta$. Of
course the size of the domains should not arbitrarily decrease
since in this case Eq.~(\ref{eq:lamdisbcr}) is not valid any more.

\section{Results for the 2D Ising model}
\label{sec:results}

We apply the method of analysis described in the previous
Sec.~\ref{sec:strategy} on the square-lattice Ising model
attempting to calculate the isothermal critical exponent $\delta$.
In the Hamiltonian of Eq.~(\ref{eq:Ising}) we use the homogeneous
case $J_{ij}=J$ for all pairs $i$ and $j$. We investigated the
equilibrium dynamics of the Ising spins, mainly for a lattice
set-up with linear size $L=256$.

In the pre-equilibrium phase the spin dynamics is simulated using
the Wolff's cluster algorithm~\cite{wolff89}, which is a variant
of the original Swendsen-Wang algorithm~\cite{swendsen87}, for
reasons that are exemplified below. In the Swendsen-Wang
algorithm, small and large clusters are created. While the
destruction of critical correlations is predominantly due to the
large clusters, a considerable amount of effort is spent on
constructing the smaller clusters. In Wolff's implementation, no
decomposition of the entire spin configuration into clusters takes
place. On the contrary, only a single cluster is formed, which is
then always flipped. If this cluster turns out to be large,
correlations are destroyed as effectively as by means of the large
clusters in the Swendsen-Wang algorithm, without the effort of
creating the smaller clusters that fill up the remainder of the
lattice. If the Wolff cluster turns out to be small, then not much
is gained, but also not much computational effort is required. As
a result, critical slowing down using the Wolff implementation is
suppressed even more strongly than in the Swendsen-Wang case.

After reaching equilibrium, a hybrid algorithm is employed
involving mainly single spin flips mixed with rare Wolff's steps
to increase ergodicity. The latter is reflected in the smooth form
obtained for the magnetization histograms. Actually the obtained
results are robust against changes of the simulation algorithm
provided that time is expressed in single spin flips. Thus, for
example, employing Wolff's cluster algorithm one has to take into
account the size of the cluster when assigning a characteristic
time to the cluster flipping. This increases the corresponding
waiting times.

In our calculations we used square domains of linear sizes $L_d=7,
13,\ldots, 27$. The pseudocritical temperature $T_{c}\approx 2.29$
is estimated through the magnetization histogram as the minimum
temperature for which symmetry breaking is not yet globally
established. In Fig.~\ref{fig:fig1}(a) - (d) we show the
magnetization probability distribution for domains with $L_d=7,
17, 27$ as well as for the entire lattice $L=256$. In each case it
is calculated using a time series of $10^6$ steps after reaching
equilibrium. In accordance with the discussion in
Sec.~\ref{sec:strategy} we clearly observe that the magnetization
distribution for decreasing linear size corresponds to a spin
system with decreasing temperature. The arrows indicate the
half-variance region around the $M=0$ value.

\begin{figure}[htbp]
\centerline{\includegraphics[width=8 cm]{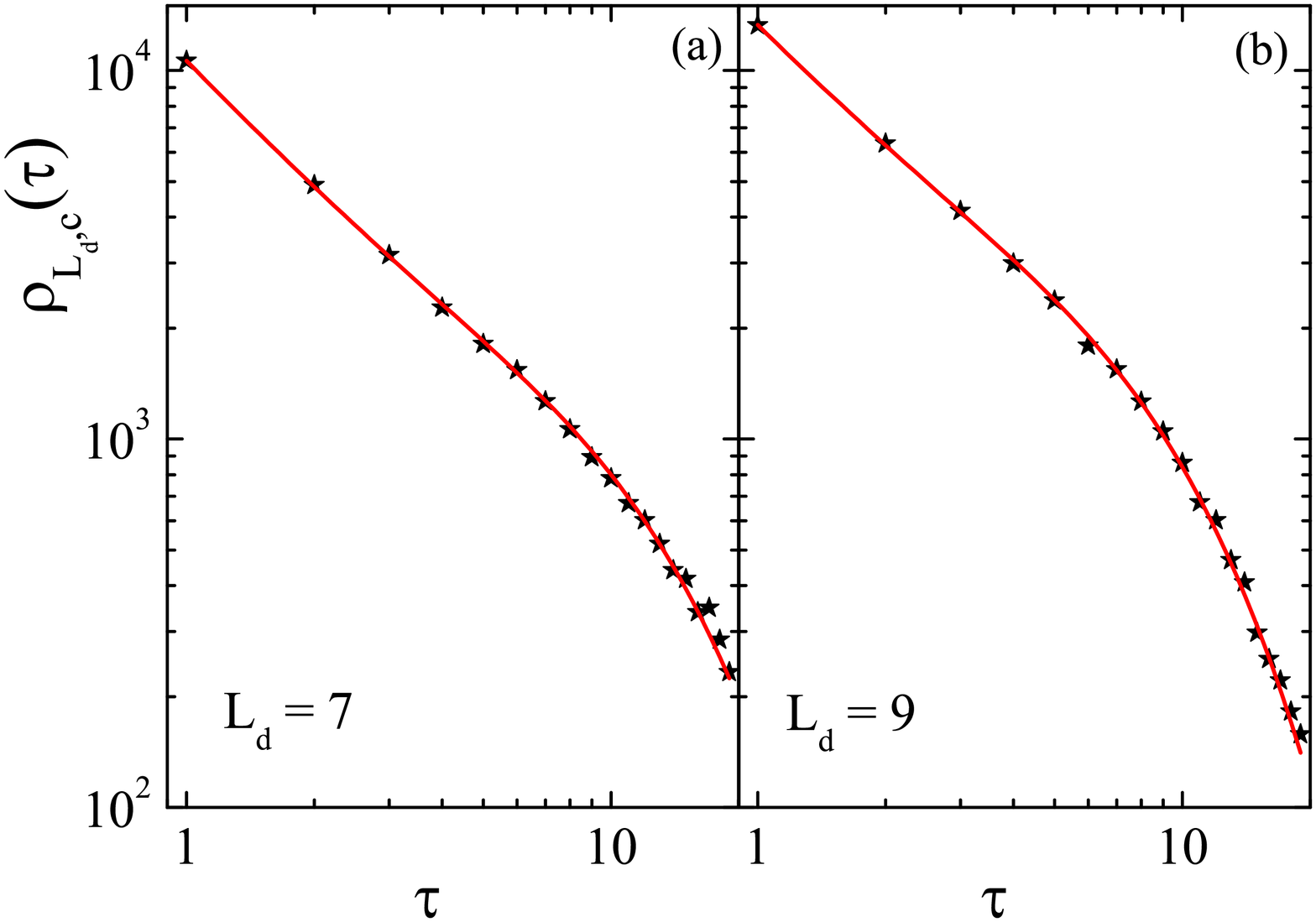}}
\caption{(color online) The distribution $\rho_{L_d,c}(\tau)$
shown by stars for (a) $L_d=7$ and (b) $L_d=9$ in a
double-logarithmic scale. In both panels the red solid line
illustrates the fitting result obtained using the function of
Eq.~(\ref{eq:lamdisbcr}).} \label{fig:fig3}
\end{figure}

In Fig.~\ref{fig:fig2}(a) - (d) we show the corresponding
distribution of the waiting times $\tau$ in the half-variance
region indicated in Fig.~\ref{fig:fig1}(a) - (d). We clearly
observe the power-law behavior for low $\tau$-values when $L_d
\geq 17$. For $L_d=7$ the effective temperature in the domain is
far below the pseudocritical one corresponding to this lattice
size [see the distribution in Fig.~\ref{fig:fig1}(a)] and
therefore no power-law behavior is clearly observed. Nevertheless
a fit using the form dictated by Eq.~(\ref{eq:lamdisbcr}) works
still sufficiently well, providing a reasonable value for the
exponent $\zeta$. This is demonstrated in Fig.~\ref{fig:fig3},
where we plot with stars the distribution $\rho_{L_d,c}(\tau)$ for
$L_d=7$ (a) and $L_d=9$ (b), respectively. The corresponding
fitting results are shown by the red solid lines in the two
panels. They lead to $\zeta(L_d) \approx 4.3$ for both values of
$L_d$. As the size of the domain decreases, the exponential tail
dominates more and more, extending also to the low-$\tau$ region
and the power-law behavior gradually disappears. Therefore we do
not consider domains with $L_d < 7$. For $L_d > 10$ we fit the
waiting-time distribution with a power law in the low-$\tau$
region ($\tau < 20$) for each considered domain and we extract the
exponent $p_2(L_d)=\zeta(L_d) / [\zeta(L_d)-1]$. We check the
validity of the power-law description by repeating the fit with
the form given in Eq.~(\ref{eq:lamdisbcr}) and comparing the
obtained value of $p_2(L_d)$ to the value found by the direct
power-law fit. We observe that for $L_d > 11$ the two values
practically coincide.

\begin{figure}[htbp]
\centerline{\includegraphics[width=8 cm]{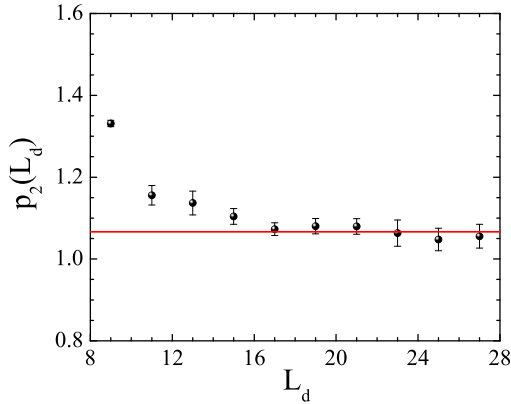}}
\caption{(color online) The exponent $p_2(L_d)$ calculated from
the distributions $\rho_{L_d}(\tau)$ as explained in the text. The
red line is at $16/15 \approx 1.067$ (analytical result).}
\label{fig:fig4}
\end{figure}

In Fig.~\ref{fig:fig4} we show graphically the results obtained
for the characteristic exponent $p_2(L_d)$ and for $L_d \geq 9$.
We observe a rapid convergence to the value $(\delta +1) / \delta
= 16/15 \approx 1.07$, shown by the red solid line. For large
domain sizes the errors increase due to the fact that the
microstate space increases exponentially and therefore the
statistical fluctuations in the calculation of the waiting time
distribution increase too. The isothermal critical exponent
$\delta$ is estimated via $\delta=\frac{1}{p_2(L_d) - 1}$. For the
2D Ising model this calculation leads to the result $\delta_{\rm
{2D}}=14.7 \pm 1.4$ (obtained by a fit with a constant for $L_d
\geq 15$) which is in very good agreement with the exact value
$\delta_{\rm {2D}} = 15$~\cite{baxter,mccoy}.

\section{Concluding remarks}
\label{sec:conclusions}

We proposed an approach for the calculation of the isothermal
critical exponent $\delta$ in a spin lattice undergoing a
second-order phase transition. Our method makes use of critical
characteristics present in post-equilibrium spin dynamics as
revealed in Refs.~\cite{Contoyiannis2002,Contoyiannis2007}. In
fact the effective magnetization dynamics close to the vacuum
undergoing the spontaneous symmetry breaking is similar to
intermittent dynamics close to a bifurcation. The information
concerning $\delta$ is encoded in the distribution of waiting
times of the magnetization time series near the broken vacuum. The
key ingredient of the method is the calculation of the
magnetization time series within domains of varying size embedded
in the spin lattice. These domains equilibrate locally to a lower
temperature than that of the entire lattice. When this temperature
is not too far from the pseudocritical one, then the waiting-time
distribution carries information of the critical point related to
$\delta$, as shown in Ref.~\cite{Contoyiannis2007}. Furthermore,
it can be calculated with higher accuracy since the number of
microstates of the smaller domain is exponentially reduced
compared to that of the entire lattice. For small to moderate
values of the waiting times, the associated distribution is a
power law with an exponent which is related to $\delta$ and can be
estimated employing a suitable fitting procedure. Subsequently the
sequence of the fitted exponents as a function of the increasing
domain size can be used for an accurate determination of $\delta$.
The proposed method has been applied to the 2D Ising model on the
square lattice providing remarkably good results. Let us note here
that our choice to use the 2D Ising ferromagnet as a platform
model was dictated by the existence of an exact analytical result
of the isothermal exponent, so that a direct comparison was
possible. However, since the dynamical properties consisting the
backbone of the method are universal, one expects that the method
can be used in a wide class of spin systems to calculate the
isothermal critical exponent in a rather straightforward manner.

\begin{acknowledgements}
P.A.~K acknowledges financial
support from the State Scholarships Foundation (IKY) Fellowships
of Excellence for Postdoctoral Research in Greece - Siemens
Programme.
\end{acknowledgements}

\end{document}